\documentclass[galaxies,review,accept,pdftex,moreauthors]{Definitions/mdpi}
\firstpage{1}
\pubvolume{11}
\issuenum{1}
\articlenumber{27}
\pubyear{2023}
\copyrightyear{2023}
\externaleditor{Academic Editors:  Stefi Baum and Christopher P. O'Dea }
\datereceived{15 December 2022}
\daterevised{21 January 2023} 
\dateaccepted{2 February 2023}
\datepublished{8 February 2023}
\hreflink{https://doi.org/10.3390/\linebreak galaxies11010027} 

\Title{Looking for Signatures of AGN Feedback in Radio-Quiet AGN}

\TitleCitation{Looking for Signatures of AGN Feedback in Radio-Quiet AGN}

\Author{{Preeti} Kharb *$^{,\dagger}$\href{https://orcid.org/0000-0003-3203-1613}{\orcidicon} and Sasikumar Silpa $^{\dagger}$\href{https://orcid.org/0000-0003-0667-7074}{\orcidicon}}

\AuthorNames{Preeti Kharb and Sasikumar Silpa}
\AuthorCitation{{Kharb}, P.; Silpa, S.}

\address[1]{%
National Centre for Radio Astrophysics{,} Tata Institute for Fundamental Research {(NCRA-TIFR),}  {\mbox{Savitribai Phule Pune University}, Ganeshkhind, Pune 411007, India} }

\corres{\hangafter=1 \hangindent=1.05em \hspace{-0.82em} Correspondence: kharb@ncra.tifr.res.in}

\firstnote{\hangafter=1 \hangindent=1.05em \hspace{-0.82em} These authors contributed equally to this work.}

\abstract{{In this article, we discuss the state of ``AGN feedback'' in radio-quiet (RQ) AGN. This study involves heterogeneous samples of nearby Seyfert and LINER galaxies as well as quasi-stellar objects (QSOs) that have been observed at low radio frequencies (few $\sim$100~MHz) with the {upgraded Giant Meterwave Radio Telescope (GMRT)} and $\sim$GHz frequencies with the {Karl G. Jansky Very Large Array (VLA)} and Very Long Baseline Array (VLBA). These multi-frequency, multi-resolution observations detect a range of arcsecond-scale radio spectral indices that are consistent with the presence of multiple contributors including starburst winds and AGN jets or winds; steep spectrum ``relic'' emission is observed as well. Polarization-sensitive data from the VLA and GMRT suggest that the radio outflows are stratified (e.g., in IIIZw2, Mrk231); distinct polarization signatures suggest that there could either be a ``spine + sheath'' structure in the radio outflow, or there could be a ``\mbox{jet + wind}'' structure. Similar nested biconical outflows can also explain the VLBA and SDSS emission-line data in the KISSR sample of double-peaked emission-line Seyfert and LINER galaxies. Furthermore, the modeling of the emission-lines  with plasma modeling codes such as MAPPINGS indicates that parsec-scale jets and winds in these sources can disturb or move the narrow-line region (NLR) gas clouds via the ``shock + precursor'' mechanism. Apart from the presence of ``relic'' emission, several Seyfert and LINER galaxies show clear morphological signatures of episodic jet activity. In one such source, NGC2639, at least four distinct episodes of jets are observed, the largest one of which was only detectable at 735~MHz with the GMRT. Additionally, a $\sim$6~kpc hole in the CO molecular gas along with a dearth of young stars in the center of its host galaxy is observed. Multiple jet episodes on the 10--100 parsec scales and a $\sim$10 parsec hole in the molecular gas is also observed in the Seyfert galaxy NGC4051. This suggests a link between episodic jet activity in RQ AGN and ``AGN feedback'' influencing the evolution of their host galaxies. However, a similar simple relationship between radio outflows and molecular gas mass is not observed in the Palomar--Green (PG) QSO sample, indicating that ``AGN feedback'' is a complex phenomenon in RQ AGN. ``AGN feedback'' must occur through the local impact of recurring multi-component outflows in RQ AGN. However, global feedback signatures on their host galaxy properties are not always readily evident.}}


\begin{document}


\section{Introduction} \label{sec:intro}
\textls[-15]{Galaxy evolution is one of the leading open questions in astrophysics (e.g., \citep{Gardner2006}). The~observational findings of a close link between the galaxy properties, {such as bulge mass and stellar velocity dispersion}, and its central supermassive black hole (SMBH) have led astronomers to believe that the SMBH with its parsec-scale sphere of influence can affect the kpc-scale galaxy bulge through the process of ``Active Galactic Nuclei (AGN) feedback'' \citep{Fabian2012, King2015}. AGN are believed to regulate galaxy growth by injecting energy into the surrounding gas, which has the effect of either heating and/or expelling star-forming gas (``negative feedback'') or facilitating localized star-formation (``positive feedback'') \citep{AlexanderHickox2012, Fabian2012, Morganti2017, Harrison2017}. ``AGN feedback'' can occur through radiative power (``quasar mode'', e.g.,~ \citep{Faucher-GiguereQuataert2012, Costa2018}) or mechanical power fed back into the galaxy through AGN winds and jets  (``maintenance/jet mode'', e.g., \citep{McNamaraNulsen2012, Mahony2013, Morganti2017, HardcastleCroston2020}). The~observational and energetic signatures of this ``AGN feedback'' are, however, still far from being unambiguous~\citep{King2015, Harrison2018}. While highly collimated jets cannot be efficient agents of ``AGN feedback'', presumably due to the smaller working surfaces at their advancing ends, relatively isotropic impacts via changes in jet direction can be highly effective, {as can broader AGN outflows and winds \citep{King2015}.} }

Radio-quiet (RQ) AGN that comprise greater than 80\% of the AGN population, however, have small-scale jets on the parsec to hundreds of parsec scales \citep{deBruyn-Wilson1976, Ulvestad-Wilson1984, Roy1994, Thean2000}, but typically not extending beyond $\sim$10~kpc \citep{Gallimore2006, Singh2015}; their total radio luminosities do not overwhelm their optical luminosities \citep{Kellermann1989}. Recent work using 6~GHz Very Large Array (VLA) data of SDSS quasi-stellar objects (QSOs) by \citet{Kellermann2016} has suggested that RQ AGN have $21 \le \mathrm{log[L_6 (W~Hz^{-1})] \le 23}$. RQ AGN also tend to be low luminosity AGN (LLAGN) and comprise Seyfert nuclei and LINER galaxies \citep{Heckman1980, Ho08}. \citet{Ho99} have identified LLAGN as those with H$\alpha$ line luminosities ranging from $10^{37}$ to $10^{41}$~erg~s$^{-1}$ or with bolometric luminosities in the range L$_{bol}\lesssim10^{37} - 10^{44}$~erg~s$^{-1}$ and Eddington ratios in the range L$_{bol}$/L$_{Edd}\sim10^{-9} - 10^{-1}$ \citep{Ho09}. In~this article, we use LLAGN and RQ AGN interchangeably when referring to Seyferts, LINERs, and~RQ quasars (which we use in lieu of QSOs).

Jet--{interstellar medium (ISM)} interaction has been inferred or observed in several RQ AGN in the literature. For~example, in IC5063, NGC5643, NGC1068, and~NGC1386 (as part of the Measuring Active Galactic Nuclei Under MUSE Microscope (MAGNUM) Survey;~\citep{Venturi21}), IC5063 \citep{Morganti2017, Tadhunter14, Dasyra2022}, NGC1266 \citep{Alatalo11}, HE  0040-1105  from the Close AGN Reference Survey (CARS) survey \citep{Singha22}, and~subset of sources from the Quasar Feedback Survey \mbox{\citep{Jarvis2019, Girdhar2022, Girdhar22b, Silpa2022}}. Evidence for an interaction between the AGN-driven winds and ISM has also been reported for RQ AGN in the literature (e.g., \citep{Feruglio10, RupkeVeilleux11, LiuG13, ZakamskaGreene14, McElroy15, WylezalekZakamska16}). Therefore, despite the lack of large and powerful radio outflows in RQ AGN, they can be effective agents for AGN~feedback.

In this review, we discuss the radio observations at multiple spatial resolutions of samples of Seyferts, LINERs, and~RQ quasars. We review the results from multi-frequency as well as multi-scale observations going from arcsecond scales with the upgraded Giant Meterwave Radio Telescope (GMRT) and {the Karl G. Jansky Very Large Array (VLA) in Sections~\ref{sec2}--\ref{sec4}, to~milli-arcsec scales with the Very Long Baseline Array (VLBA),} of nearby low luminosity or radio-quiet AGN in Section~\ref{sec5}, and~we discuss the state of ``AGN feedback'' in them. {All sources lie at redshifts $<$0.1, which correspond to spatial scales smaller than $\sim$2~parsec on milli-arcsec and $\sim$2~kpc on arsec scales.} We assume a cosmology with \mbox{H$_0$ = 73~km~s$^{-1}$}~Mpc$^{-1}$, $\Omega_{mat}$ = 0.27, $\Omega_{vac}$ =  0.73. Spectral index $\alpha$ is defined such that the flux density at frequency $\nu$ is $S_\nu \propto \nu^{\alpha}$.

\section{Low Radio Frequency Observations of RQ~AGN}\label{sec2}
Low-frequency radio (few $\sim$100~MHz) observations with the GMRT have typically detected radio emissions from both the AGN as well as stellar-related activity from the host galaxies of Seyferts or LINERs (e.g., \citep{Singh2013, Singh2015, Hota06, Kharb16}). In~these data, radio emission from the Seyferts/LINERs is lobe-like or bubble-like but often simply in the form of extended diffuse emission without distinct morphological features. The~radio spectral index images cannot distinguish between the AGN and stellar-related emission either. While the stellar-related emission is typically steep ($\alpha \le-0.7$), the AGN emission even in the unresolved ``cores'' can be either flat ($\alpha \ge-0.5$) or steep. { Steep spectrum ``cores'' are likely to be due to the large beam of the GMRT including optically thin jet or lobe emission on sub-arcsec scales.} The presence of additional lobes {on arcsec scales has} been indicated in the GMRT 325 and 610 MHz observations of the Seyfert galaxies, NGC3516, NGC5506, NGC5548, NGC5695 (Rubinur~et~al. in prep.).

\begin{figure}[H]
\begin{adjustwidth}{-\extralength}{0cm}
\centering
\includegraphics[width=17cm]{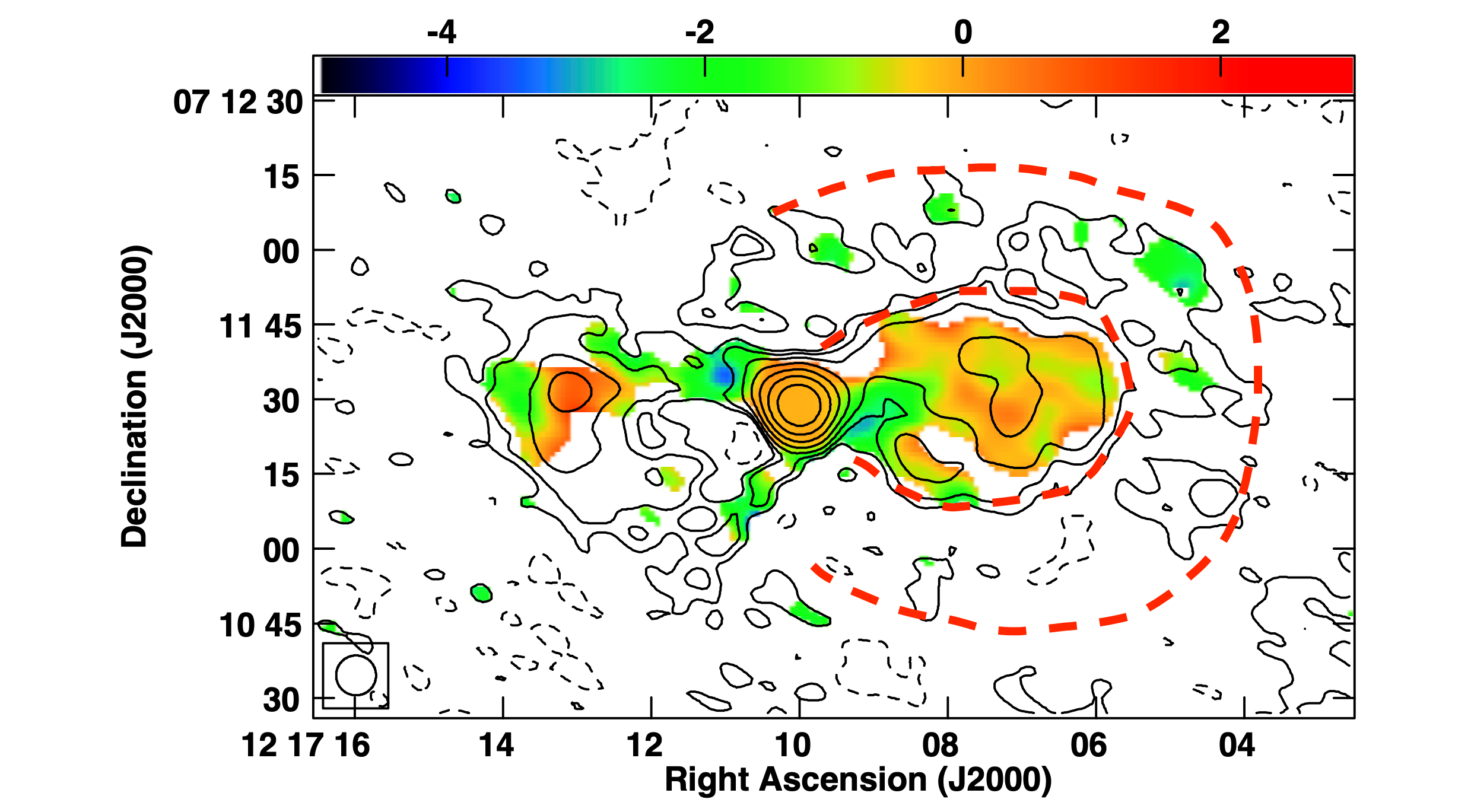}\end{adjustwidth}
\caption{The  325{--}610~MHz spectral index image from the GMRT in color overlaid with the 610~MHz radio contours for the Seyfert galaxy NGC4235. The~average spectral index is $\sim$$-$0.6 in the lobes and $\le$$-$1.8 in the ``relic'' lobe ({emission between the red dotted lines}). The~contour levels are in percentage of the peak intensity and increase {by factors of two}. The~peak intensity and lowest contour levels are $2.3\times10^{-2}$~Jy~beam$^{-1}$ and $\pm$0.3\% {of the peak intensity}, respectively. The~contour image is convolved with a circular beam of size 8~arcsec. Image reproduced from \citet{Kharb16}.}
\label{fig1}
\end{figure}

GMRT observations can probe additional activity episodes as large steep-spectrum diffuse lobes/bubbles or as ``relic'' lobe emission that is no longer being supplied with particles and fields from the AGN (e.g., \citep{Kharb16, Silpa2020, Silpa2021a, Rao2023}). The~lobes of the Seyfert galaxy NGC4235 with abruptly changing spectral indices between two sets of lobes {represent} one such case (see Figure~\ref{fig1}; \citep{Kharb16}). Both the 610~MHz image and the 325--610 MHz spectral index image of NGC4235 hinted at diffuse steep-spectrum radio emission just beyond the well-defined western lobe and enveloping it. While the average spectral index is $-0.29\pm0.19$ in the western lobe and $-0.56\pm0.24$ in the eastern lobe, the~average spectral index is $-1.82\pm0.17$ in this extended region. {The average spectral index errors come from a spectral index noise image. The~robustness of the steep spectrum emission is discussed in greater detail by \citet{Kharb16}.} This surrounding steep-spectrum emission is reminiscent of ``relic'' radio emission, as~observed in the lobes of the radio galaxy 3C388 by \citet{Roettiger94}, where a lobe with an average spectral index of $\sim$$-$0.8 was surrounded by the steep-spectrum emission of spectral index $\sim$$-$1.5 from a previous AGN activity episode. Based on a simple spectral aging analysis in NGC4235, the~relic outer lobe appeared to be at least two times older than the present lobe. This implied that the AGN in NGC4235 was switched ``off'' for the same time that it has been ``on'' for the current episode \citep{Kharb16}.
\vspace{-10pt}

\section{Jet Driven Feedback: The Case of~NGC2639}\label{sec3}
\citet{Sanders1984} had predicted that Seyfert activity must be episodic on timescales of \mbox{$10^4$--$10^5$~years} during its statistical lifetime of  3--7  $\times$ $10^8$~years. They had arrived at these numbers based on the typical extent of the radio outflows as well as the NLR in Seyfert galaxies. A~Seyfert galaxy must therefore undergo $\sim$100  activity episodes during its lifetime. Clear examples of episodic jet activity in Seyfert and LINER galaxies have been identified in only a couple of sources, viz., Mrk6 \citep{Kharb2006}, NGC2992 \citep{Irwin2016}, and~NGC2639 \citep{Sebastian2019}. \mbox{\citet{Sebastian2020}} had found tentative signatures for multiple jet episodes in a majority ($\sim$55\%) of their small Seyfert galaxy sample of nine sources. This fraction appeared to be higher than that has been reported in radio-loud (RL) AGN ($\sim$10--15\%  in radio galaxies~\citep{Jurlin2020}) {(see also the 3D GRMHD jet simulations of \citet{Lalakos2022} that reproduce these low fractions)}. However, signatures of $\sim$100 episodes are almost never observed in Seyferts or LINERs, at~least in the radio outflows. This might not be due to their true absence but rather due to the difficulty in their identification due to the low surface brightness of Seyfert/LINER lobes, their small spatial extents, lack of collimated jets, and~confusion with the radio emission from stellar-related activity (star-formation, supernovae, and~starburst winds) arising in the host~galaxy.

\begin{figure}[H]

\begin{adjustwidth}{-\extralength}{0cm}
\centering 
\includegraphics[width=17cm]{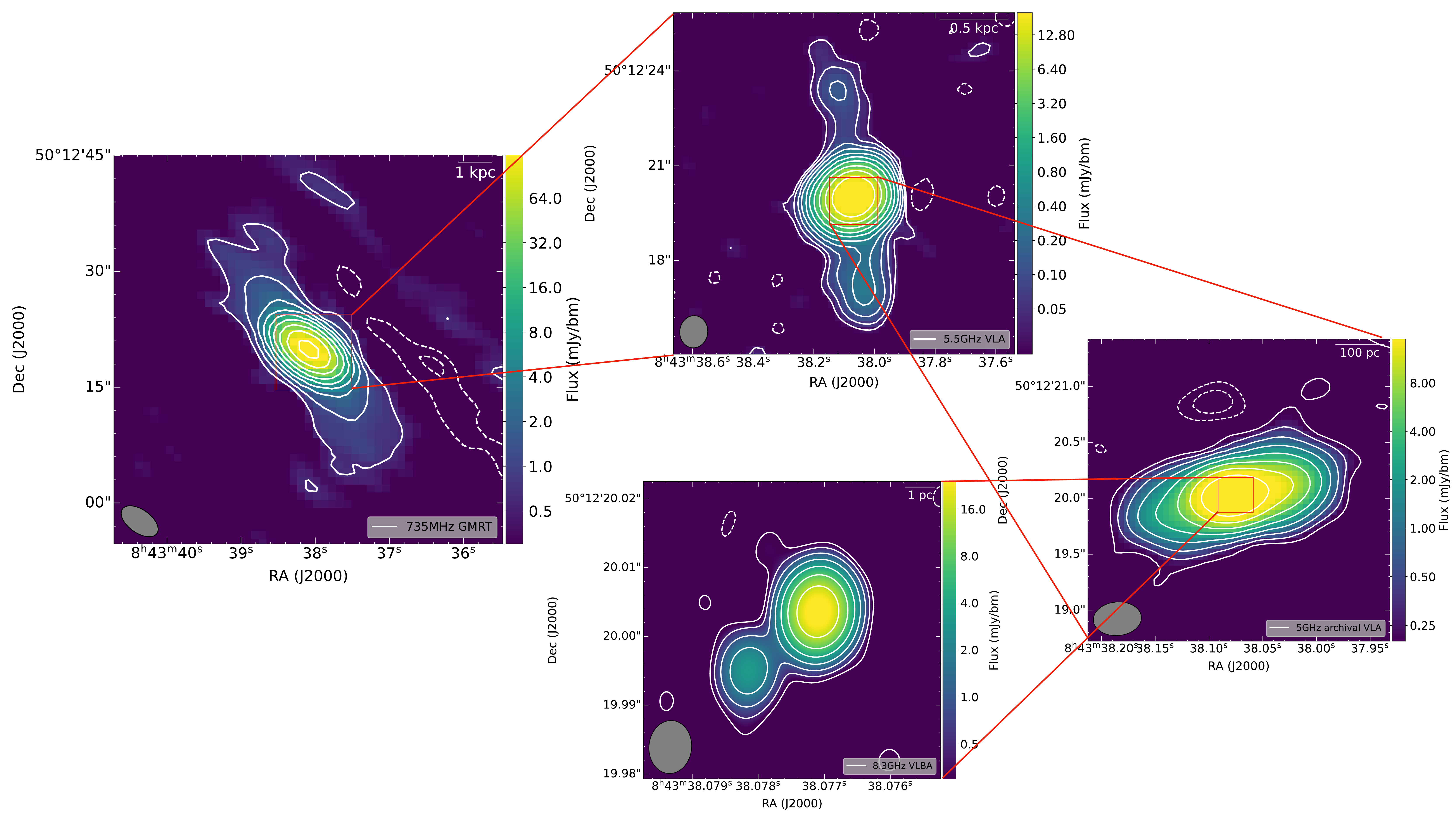}
\end{adjustwidth}
\caption{The four AGN jet episodes of NGC2639. (Left) 735~MHz GMRT total intensity image. The~$\sim$9~kpc radio lobes are seen in this image. Contour levels: $(-2, -1, 1, 2, 4, 8, 16, 32, 64, 128, 256)\times0.6$~mJy~beam$^{-1}$. The~beam at the bottom left corner is of size: 5.48~arcsec $\times$ 3.0~arcsec at PA = $54.6$~degrees. (Top) 5.5~GHz VLA total intensity image. Contour levels: $(-2, -1, 1, 2, 4, 8, 16, 32, 64, 128, 256, 512)\times 0.03$~mJy~beam$^{-1}$. The~$\sim$1.5~kpc north-south radio jets are seen here. Beam size: 1.02~arcsec $\times$ 0.89~arcsec at PA = $-5.8$~degrees. (Right) 5~GHz VLA radio image. Contour levels: $(-2, -1, 1, 2, 4, 8, 16, 32, 64, 128)\times 0.164$~mJy~beam$^{-1}$. The~$\sim$360 parsec east-west lobes are seen in this image. Beam size: 0.43~arcsec $\times$ 0.30~arcsec at PA = $-85.4$~degrees. (Bottom) 8.3 GHz VLBA image showing a $\sim$3~parsec jet at PA = 130~degrees. Contour levels: $(-2, -1, 1, 2, 4, 8, 16, 32, 64)\times0.239$~mJy~beam$^{-1}$. Beam size: 7.7~mas $\times$ 6.2~mas at PA = $-$4.9~degrees. Figure reproduced from \citet{Rao2023}. {A $\sim$6~kpc hole in the molecular gas distribution is observed in the central regions of NGC2639.}}
\label{fig2}
\end{figure}

\citet{Sebastian2019} had noted the presence of at least three episodes in NGC2639 with two sets of bipolar radio lobes detected with the VLA and oriented nearly perpendicular to each other, similar to what was observed in Mrk6, as~well as a parsec-scale core--jet structure detected with the Very Long Baseline Array (VLBA) and oriented at least 30~degrees from the sub-kpc east--west oriented radio lobes. GMRT observations at 325 and 735 MHz showed the presence of an additional pair of radio lobes, oriented nearly 45~degrees from the previously known north--south lobes (see Figure~\ref{fig2}; \citep{Rao2023}). {It is worth pointing out that there is no continuous connecting radio emission between different jet episodes. Rather, the~jets and lobes in each episode have clearly defined hotspots or edges distinguishing them as independent events, making a single precessing jet model fitting all of the radio emission improbable.}

Ages of the three pairs of lobes were derived using the spectral aging software BRATS, which stands for
Broadband Radio Astronomy Tools \citep{Harwood2015}, and they turned out to be, respectively, $34^{+4}_{-6}$ million years, 
$11.8^{+1.7}_{-1.4}$ million years, and~$2.8^{+0.7}_{-0.5}$ million years, with~the GMRT lobes being the oldest. Using the ``on'' and ``off'' times of these {jets or lobes (using the spectral age of a given set of lobes and the spectral age difference between two sets of lobes, respectively)}, the~AGN jet duty cycle in NGC2639 turns out to be $\sim$60\%. Based on the molecular gas data from the EDGE \citep{Bolatto2017}---Calar Alto Legacy Integral Field Area (CALIFA~\citep{Sanchez2012}) survey, \citet{Ellison2021} have found that the gas fraction in the central region of NGC2639 is a factor of a few lower than in star-forming regions, suggesting that the AGN has partially depleted the central molecular gas reservoir. Like the CO (1-0) molecular gas image, which shows a hole with a diameter of $\sim$6~kpc, the~GALEX NUV image also shows a deficiency of star formation in the last 200 million years in the inner $\sim$6~kpc region of NGC2639. These results point to star-formation quenching taking place in the central regions of NGC2639.

If the CO (1-0) molecular gas ring is a result of push-back from the jet in NGC2639, the~$PV$ (pressure times volume) amount of work done on the molecular gas by the jet to create a cavity can be estimated.  {\citet{Rao2023} have shown that} for the CO gas ring radius of 3~kpc, the~volume of the disk-like cavity is $1.25\times10^{66}$~cm$^3$, and the $PV$ work done is \mbox{$>$3.44 $ \times$ $ 10^{54}$~erg.} Using the lobe flux densities at 5~GHz, the~time-averaged power (e.g., using the relations in \citep{Merloni2007}) for a spectral age of 2.8 million years is $6.8\times10^{56}$~erg. Therefore, only $\sim$0.5\% of the east--west jet power is sufficient to push back the CO gas in NGC2639. Similarly small fractions are needed from the north--south jets and the north--east--south--west jets. However, the~creation of a hole in the molecular gas in the galactic center likely required several jet episodes to occur, given that each jet episode is collimated and therefore highly directional \citep{Rao2023}.

\begin{figure}[H]

\begin{adjustwidth}{-\extralength}{0cm}
\centering 
\includegraphics[width=17cm]{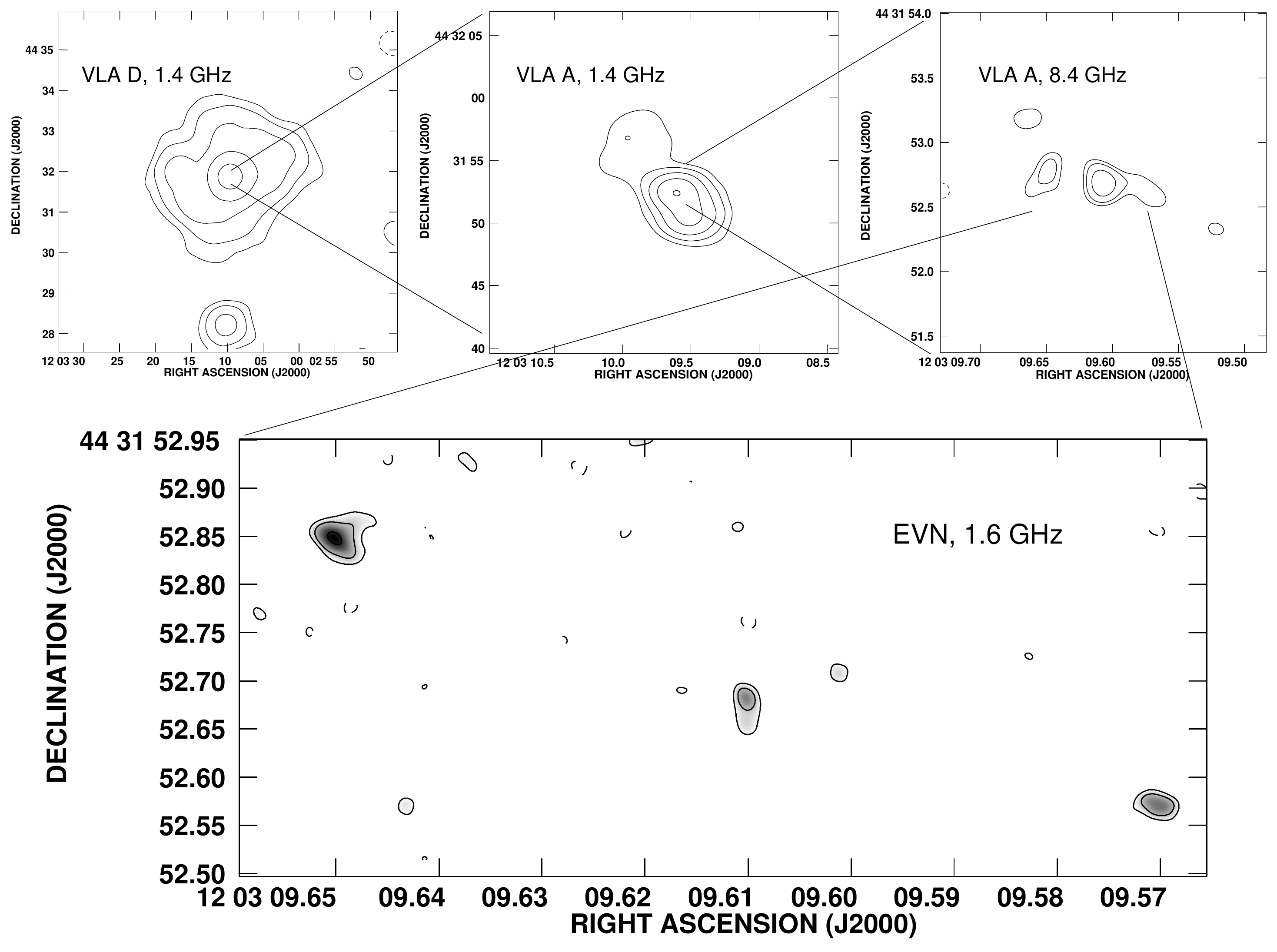}
\end{adjustwidth}
\caption{The three radio lobes or jets at different position angles in the Seyfert galaxy NGC4051. The~leftmost panel (VLA 1.4 GHz) image shows the galactic emission surrounding the 1~kpc-sized radio AGN. The~middle panel shows $\sim$830~parsec lobes at a PA of 30 degrees. The~rightmost panel shows a $\sim$90~parsec lobes at a PA of 60 degrees. The~bottom panel shows the EVN image with a {parsec-scale core--jet structure, showing a faint extension toward the south.} Figure reproduced from \citet{Giroletti2009}. {A $\sim$10~parsec hole in the molecular gas distribution is observed in the central regions of NGC4051.}}
\label{fig3}
\end{figure}

NGC2639 or Mrk6 are not unique in showing signatures of radio {jets or lobes} that have different sky orientations when looked at with multiple spatial resolutions. Another case is the famous Seyfert galaxy, NGC4051. Multi-resolution images of NGC4051 from \citet{Jones2011} and \citet{Giroletti2009} show at least three sets of radio lobes or jets. At~a resolution of $\sim$2 arcsec, there lies a double-lobed radio structure of extent $\sim$830~parsec at a PA of 30 degrees (at its redshift of 0.002336, 1 arcsec corresponds to 61 parsecs). At~a resolution of $\sim$0.5 arcsec, there is a jet--core--jet structure of extent $\sim$90~parsec at a PA of 60~degrees. Finally, on the 100 mas scale as probed by {the European VLBI Network (EVN), the~parsec-scale core--jet structure shows a faint extension toward the south} (See Figure~\ref{fig3}). Interestingly, NGC4051 also appears to have a hole with a diameter of $\sim$10~parsecs in its molecular gas distribution, as observed from Gemini North (data from \citet{Riffel2008} reanalyzed by D. May~et~al. \endnote{\url{https://webarchive.gemini.edu/20210427-gsm12/gsm2012_poster_may-daniel.pdf}~(accessed~on~1~January~2023).}). Overall, NGC2639 and NGC4051 could be candidates for the presence of jet-driven ``negative AGN feedback'' in RQ~AGN.

\section{Jet + Wind Driven~Feedback}\label{sec4}
\unskip
\subsection{The Palomar--Green Radio-Quiet Quasar Study on~Kpc-Scales}\label{sec41}
\begin{figure}[H]


\begin{adjustwidth}{-\extralength}{0cm}
\centering 
{\includegraphics[width=18cm]{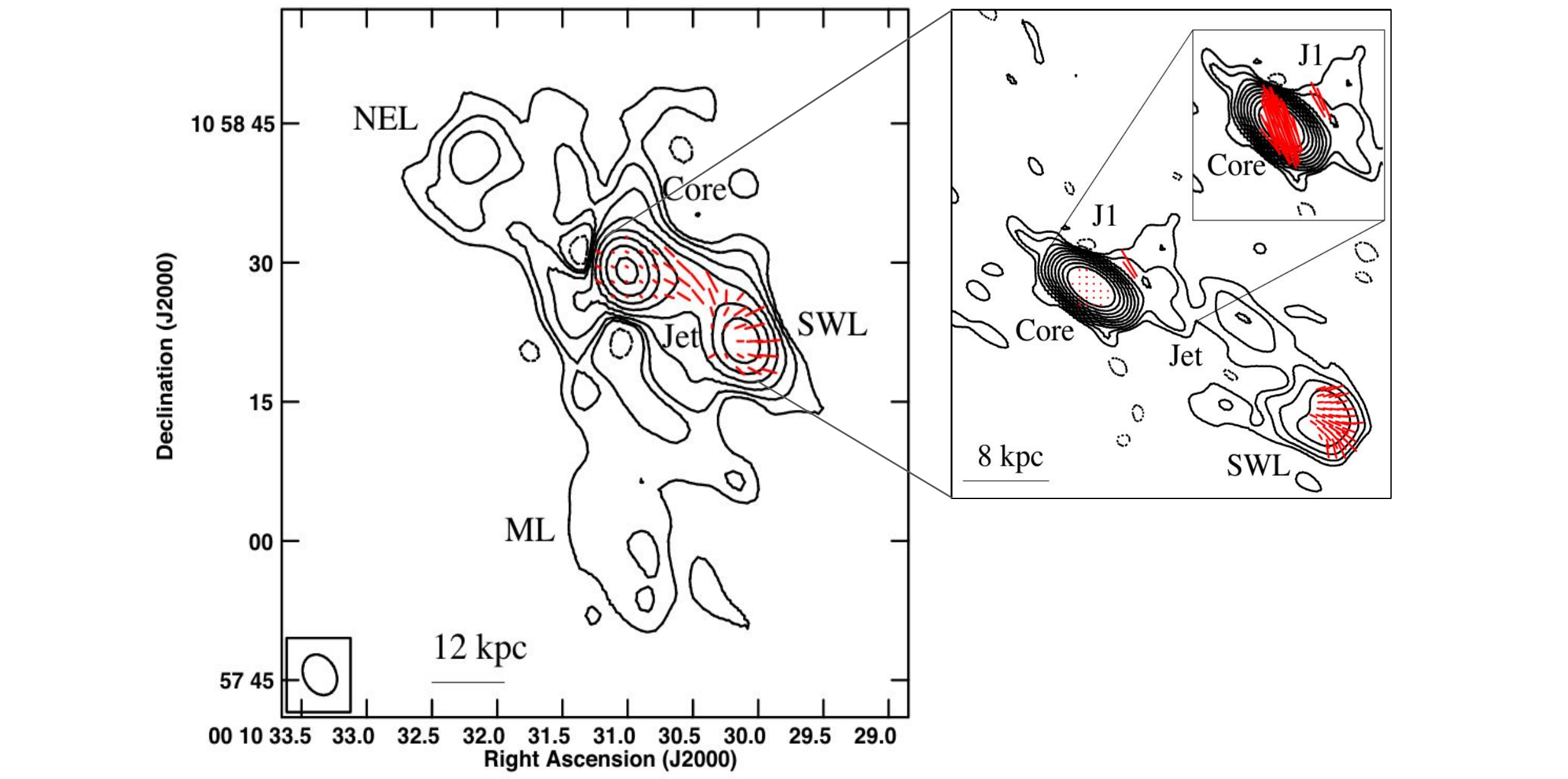}}
\end{adjustwidth}
\caption{The GMRT 685~MHz total intensity contour image of IIIZw2 superimposed with red fractional polarization vectors, with 2~arcsec length of the vector corresponding to 6.25\% fractional polarization. {The inset on the right presents the VLA 5~GHz total intensity contour image of the south-western jet--lobe region, superimposed with red fractional polarization vectors, with 1~arcsec length of the vector corresponding to 25\% fractional polarization. The~smallest inset on the top right presents the VLA 5~GHz total intensity contour image of the radio core with red polarized intensity vectors, with 1 arcsec length of the vector corresponding to 0.031 mJy beam$^{-1}$ polarized intensity. The~peak contour surface brightness is {\it x} mJy beam$^{-1}$ and the levels are {\it y} $\times$ ($-$1, 1, 2, 4, 8, 16, 32, 64, 128, 256, 512) mJy beam$^{-1}$, where {\it (x ; y)} are (35, 0.21) and (124, 0.05) for the GMRT and VLA images respectively.} Figure reproduced from \citet{Silpa2021a}.}
\label{fig4}
\end{figure}

{We now discuss the case of the Palomar--Green (PG; \citep{BorosonGreen92}) sample of RQ quasars, which has extensive multi-resolution radio jet and multi-phase gas outflow data, in~the literature.} \citet{Silpa2020} carried out a 685~MHz GMRT study of the PG RQ quasar sample and found that the two-frequency radio spectral indices (using the GMRT and VLA) were ultra-steep ($\alpha \le -1.0$) in few of the sources. Other than a correlation between the {total} GMRT 685~MHz luminosity and Eddington ratios, other radio properties of the sample {such as radio core sizes and radio spectral indices} did not correlate with BH properties such as their masses or Eddington ratios. This suggested either that the radio emission was stellar-related or was due to previous jet episodes (i.e., ``relic'' emission) in them (see Section~\ref{sec3}). A~combined GMRT--VLA study of the PG RQ sample has shown evidence for the presence of small-scale, bent, and~low-powered jets in a couple of sources (see Section~\ref{sec5} ahead), while for the rest, it was difficult to {unambiguously determine} the origin of radio emission (whether via jet or wind or stellar processes; \citep{Silpa2023}). Multi-frequency, multi-resolution radio polarization observations have detected a stratified radio outflow in PG 0007+106 (a.k.a. IIIZw2 \citep{Silpa2021a}). The~stratified radio outflow could either be a ``\mbox{spine + sheath}'' structure in the jet or a ``\mbox{jet + wind}'' composite structure (see Figure~\ref{fig4}). {Each component of the stratified outflow is observed to have a characteristic magnetic (B) field geometry (e.g., \citep{MehdipourCostantini19, Miller12}). B fields are aligned with the jet direction in the case of a jet or jet ``spine'' while they are transverse in the case of a wind or jet ``sheath''. The~parallel B fields could represent the poloidal component of a large-scale helical B field, while the transverse B fields could either represent the toroidal component of the helical field or a series of transverse shocks that order B fields by compression. Alternately, they could represent toroidal B fields threading an AGN wind or jet ``sheath'', which is sampled in the lower-resolution images.} The bow-shock-like feature at the termination point of the VLA jet {(see Figure~5 of \citep{Silpa2021a})}, and~the presence of a misaligned ``sputtering'' lobe in the GMRT image (annotated as ML in Figure~\ref{fig4}), are consistent with restarted jet activity in~IIIZw2.

The 5~GHz jet kinetic power of a subset of the PG RQ quasars estimated using the empirical relation of \citet{Merloni2007} {is within the range} 10$^{42}$--$10^{43}$~erg~s$^{-1}$ \citep{Silpa2023}. For~these jet powers, either stellar-mass loading (e.g.,  \citep{Komissarov1994, Bowman1996}) or the growth of Kelvin--Helmholtz (KH) instabilities triggered by recollimation shocks/jet--medium interactions can effectively decelerate and decollimate the jets (e.g.,  \citep{PeruchoMarti2007, Perucho2014}). As~the jets propagate through the host galaxy, they would encounter numerous stars, which inject matter into the jets via stellar winds. The~mixing of the injected stellar material with the jet plasma causes the jets to decelerate. {Using a small sample of RQ quasars, \citet{Silpa2022} found that the polarized radio emission and [O~III] emission did not always spatially overlap. This was suggested to arise from the depolarization of radio emission by either an irregular Faraday screen of clumpy emission-line gas or by the emission-line gas that had entrained and mixed with the synchrotron plasma in the lobes. While modeling the former scenario, the~fluctuation scales of the electron density (or, equivalently the sizes of the emission-line gas clumps or clouds, or~filaments) were considered and estimated. This value was also assumed to represent the fluctuation scales of the random B-field component while modeling the latter scenario.} Interestingly, {the lower limit on the size} value ($\sim$$10^{-5}$~parsec) matches the sizes of red giant stars. KH instabilities also promote entrainment and mixing between the surrounding gas and the jet plasma, resulting in the deceleration and decollimation of the jets. KH instabilities can also cause jet bending (e.g.,  \citep{Hardee1987, Savolainen2006}). Signatures of KH instabilities include knotty polarization structures and the presence of poloidal magnetic fields (e.g.,~\citep{Mukherjee2020}). {Both of these signatures are observed in the few RQ quasars that are detected in polarized light.} {\it {Overall}, therefore, there is evidence for small (arcsec) scale jet--medium interaction taking place in these sources.}

\citet{Shangguan2020} have found that the molecular gas masses and kinematics of the PG quasars were similar to those of the star-forming galaxies. Additionally, no molecular gas outflows were detected in these sources. Their host galaxies were found to be following the ``Kennicutt--Schmidt law'' \citep{Kennicutt1998}, suggesting that no star formation quenching was taking place in them. In~the recent work of \citet{Molina2022a}, the~molecular gas kinematics of the PG quasar host galaxies also suggest that the ``negative AGN feedback'' is ineffective in them. { In general, there appears to be no immediate significant impact on the global molecular gas reservoirs by jets or outflows in the PG sources. However, the~possibility that the AGN might be impacting the ISM locally cannot be ruled out (e.g.,  \citep{Jarvis2020}). As~discussed above, we infer the presence of localized jet--medium interaction from the jet kinetic power argument as well as the polarization data in~the PG RQ quasar sample. Localized impact on the gas by AGN or jets could in principle be captured by spatially resolved observations of multiple and higher CO transitions (e.g.,  \citep{vanderWerf2010, Mashian2015, Carniani2019, Dasyra2016, Zhang2019, Salome2017, Rosario2018, Fotopoulou2019, Ramakrishnan2019, Shin2019, Lutz2020}).}

One of the plausible mechanisms for star-formation quenching as proposed in the literature includes galaxy major and minor mergers (along with ``AGN feedback'' in most cases~\citep{DiMatteo05, Springel05a, Cox06, Hopkins08, Khalatyan08, Cheung12, Barro13, Smethurst15}). The~galaxy major merger simulations by \citet{Springel05b} show that the presence of accreting BHs can significantly impact the merger dynamics. Galaxy mergers can also cause enhanced star-forming activity \citep{SandersMirabel96, MihosHernquist96, Lin07} just as a ``positive AGN feedback'' process \citep{Mahoro17, Mahoro19}. On~the other hand, \citet{Weigel17} suggest that the {major merger quenching cannot fully explain the slow evolution of galaxies from blue to red;} alternative quenching mechanisms (``AGN feedback'' being a potential candidate) are needed. The~host galaxies of a large fraction of the PG RQ quasars show signatures of ongoing galactic mergers \citep{Silpa2020}. Therefore, the~merger scenario can also have an influence on the overall interpretation of AGN~feedback.

\subsection{The Jet and Wind in~Mrk231}\label{sec231}
{We now discuss the case of the quintessential AGN that clearly hosts both a jet and wind component and is routinely considered in ``AGN feedback'' scenarios, viz., Mrk231.} Mrk231 is a Seyfert galaxy that hosts multi-phase multi-scale gas outflows, such as a nuclear ultra-fast outflow (UFO; \citep{Feruglio2015}), a~sub-kpc scale HI outflow \citep{Morganti2016}, a~kpc-scale molecular CO outflow \citep{Feruglio10, Feruglio2015}, and a $\sim$3~kpc scale outflow seen in the Na~I doublet lines \citep{RupkeVeilleux11, RupkeVeilleux2013}. A~massive molecular OH outflow has also been detected in Mrk231 \citep{Fischer2010, Sturm2011, Morganti2017}. In~VLA observations, Mrk231 reveals a one-sided radio outflow to the south, comprising a weakly collimated jet or jet ``spine'' embedded inside a broader magnetized wind component~\citep{Silpa2021b}, resembling the stratified radio outflow in IIIZw2 (see Section~\ref{sec41}). The~composite outflow in Mrk231 is curved, low-powered, and~oriented at a small angle to our line of sight. The~wind component may comprise both a nuclear starburst and AGN wind, where the former may be the primary contributor close to the core, while the latter may dominate further away. Moving away from the core, the~wind component could also be the outer layers of a widened jet like a jet ``sheath''. The~10-kpc-scale radio structure in Mrk231 is ``self-similar'' to the radio structure observed on the 10-parsec-scale in the literature (see Figure~1 in \citep{Morganti2016}), resembling the lobes observed in Mrk6 \citep{Kharb2006}. The~radio structures on the two scales in Mrk231 are not as clearly delineated, however, which {may be a result of the low inclination angle.} However, the~presence of two distinct structures is consistent with episodic jet activity in Mrk231, similar to the case of Mrk6 (see Section~\ref{sec3}).

\citet{Silpa2021b} obtained first-order estimates of the relative contributions of the different components (jet/AGN wind/starburst wind) to the overall budget of the radio emission. While the starburst-driven wind accounted for $\sim$10--20\% of the total radio emission, both jet and AGN wind contributed significantly to the rest of the emission. To~estimate the contribution of starburst-driven wind, two different models were used: one assumed that 10\% of the supernova kinetic energy was carried away by the winds \citep{Thornton98, Leitherer99}, and~the other assumed 40\% \citep{Dalla-VecchiaSchaye08}. The~latter analysis has also been carried out assuming two different initial mass functions (IMFs; \citep{Salpeter55, Chabrier03}). The~contribution of the starburst-driven wind remained the same for the different assumed models. The~contribution of AGN wind was estimated assuming two different coupling efficiencies (5\%; \citep{DiMatteo05}) and (0.5\%;~\citep{HopkinsElvis10}). For~a 0.5\% coupling efficiency, the~radio contribution of the jet estimated using \citet{Leitherer99} model, as~well as \citet{Dalla-VecchiaSchaye08} model with \citet{Chabrier03} IMF, turned out to be more than of the wind. {Although such a first-order analysis cannot provide a one-to-one correspondence between the dominant driver (between jet or AGN wind) and the outflowing gas phase (HI or Na I D or CO or OH), it still indicates that the multi-phase gas outflow in Mrk231 is likely to be driven by both a jet and a wind.}

\section{Signatures of Jet + Wind Feedback on~Parsec-Scales}\label{sec5}
\textls[-15]{We now turn our attention to much smaller spatial scales than probed by the GMRT and VLA observations. {We focus on the KPNO Internal Spectroscopic Survey Red (KISSR;~\citep{Wegner03}) sample of Seyfert and LINER galaxies \citep{Kharb2021}. This sample was chosen based on the presence of double-peaked emission lines in their SDSS spectra as well as a radio detection in the VLA FIRST survey \citep{Kharb2015}.} Phase-referenced observations of this sample with the Very Long Baseline Array (VLBA) have revealed the presence of elongated jet-like features, similar to other Very Long Baseline Interferometry (VLBI) studies of RQ AGN in the literature (e.g.,  \citep{Falcke00, Middelberg04, Nagar05, Kharb2010, Orienti10, Baldi2018, Kharb2021}). Interestingly, these jets are one-sided, similar to the one-sided parsec-scale jets observed in RL AGN. {VLBI imaging also reveals one-sided jets in narrow-line Seyfert 1 galaxies \citep{Doi2011, Richards2015A, Richards2015B}.} A $\sim$60~parsec long radio source (jet--core--counterjet) was imaged in the Seyfert 2 galaxy Mrk348 using VLBI by~\mbox{\citet{Neff1983}}; the counterjet emission was detected more than 40~parsecs away from the radio core. In~RL AGN, one-sidedness is typically understood to be a consequence of Doppler-boosting effects \citep{Blandford1979}. It is not clear if the one-sidedness in Seyfert or LINER jets is a result of Doppler-boosting effects or due to free-free absorption; {multi-frequency and spectral index observations could help to disentangle these two scenarios (e.g.,  \citep{Baczko2019, Baczko2022}).} If Doppler-boosting is responsible for the jet one-sidedness in the KISSR sources, for instance, then lower limits to jet speeds would range from $0.003c$ to $0.75c$ \citep{Kharb2021} assuming jet inclinations to be $\ge$50~degrees, consistent with their type 2 classification, and~the expected torus half-opening angles being $\sim$50~degrees (e.g.,  \citep{Simpson96}). }

On the other hand, if~the missing counterjet emission was a result of free--free absorption, the~required electron densities of the ionized gas could be estimated using $EM = 3.05\times10^6~\tau~T^{1.35}~\nu^{2.1}$, and~$n_e=\sqrt{EM/l}$ \citep{Mezger67}, where $EM$ is the emission measure in pc~cm$^{-6}$, $\tau$ is the optical depth at frequency $\nu$ in GHz, $T$ {is} the gas temperature in units of $10^4$~K, $n_e$ {is} the electron density in cm$^{-3}$, and $l$ {is} the path length in parsecs. In~order to account for the observed jet-to-counterjet surface brightness ratios ({$R_J$}) of $\sim$20 on parsec scales (as in the case of KISSR434; \citep{Kharb2019}), the~optical depth at 1.5~GHz would need to be at least $\sim$1.0 using $\exp(-\tau)=1/R_J$ \citep{Ulvestad99}. For~a gas temperature of $10^4$~K and a jet path length of 1~parsec, an $EM$ of $\approx$7.1 $\times$ $10^6$~pc~cm$^{-6}$ and $n_e$ of $\approx$2700~cm$^{-3}$ are required for free--free absorption on parsec scales. Such ionized gas densities can be found in NLR gas clouds. However, their volume filling factor is of the order of $10^{-4}$ (e.g.,  \citep{Alexander99}), making them unlikely candidates for absorbers for the $\sim$100-parsec-scale (counter-)jets observed in a couple of the KISSR sources \citep{Kharb2019, Kharb2021}. Ionized gas in giant HII regions with \mbox{$n_e\sim$100--1000~cm$^{-3}$} could in principle also be the candidate media for free--free absorption~\mbox{\citep[][]{Clemens10}} but also has a low volume filling factor ($\ge$0.2, e.g., \citep{Walterbos94}).

Multi-epoch VLBI observations are therefore necessary to check for proper motions and obtain accurate jet speeds { to verify the Doppler-boosting picture in Seyferts and LINERs. This has recently become possible for eight KISSR sources. Preliminary results suggest the detection of superluminal jet motion in at least one source (Kharb~et~al. in prep.). With~these new observations, the~jet detection rate in the KISSR sample becomes $\sim$75\%.} Jet detection rates in larger samples of Seyferts and LINERs range between $\sim$30\% and $\sim$50\% (e.g.,  \citep{Baldi2018, Baldi2021}). The~higher jet detection rate in the KISSR sample is consistent with a selection bias that comes from selecting {double-peaked emission-line AGN (DPAGN)} and jet--NLR interaction being the primary cause of the double-peaked emission lines in~them.

\begin{figure}[H]

\hspace{-3cm}{\includegraphics[width=13.5cm]{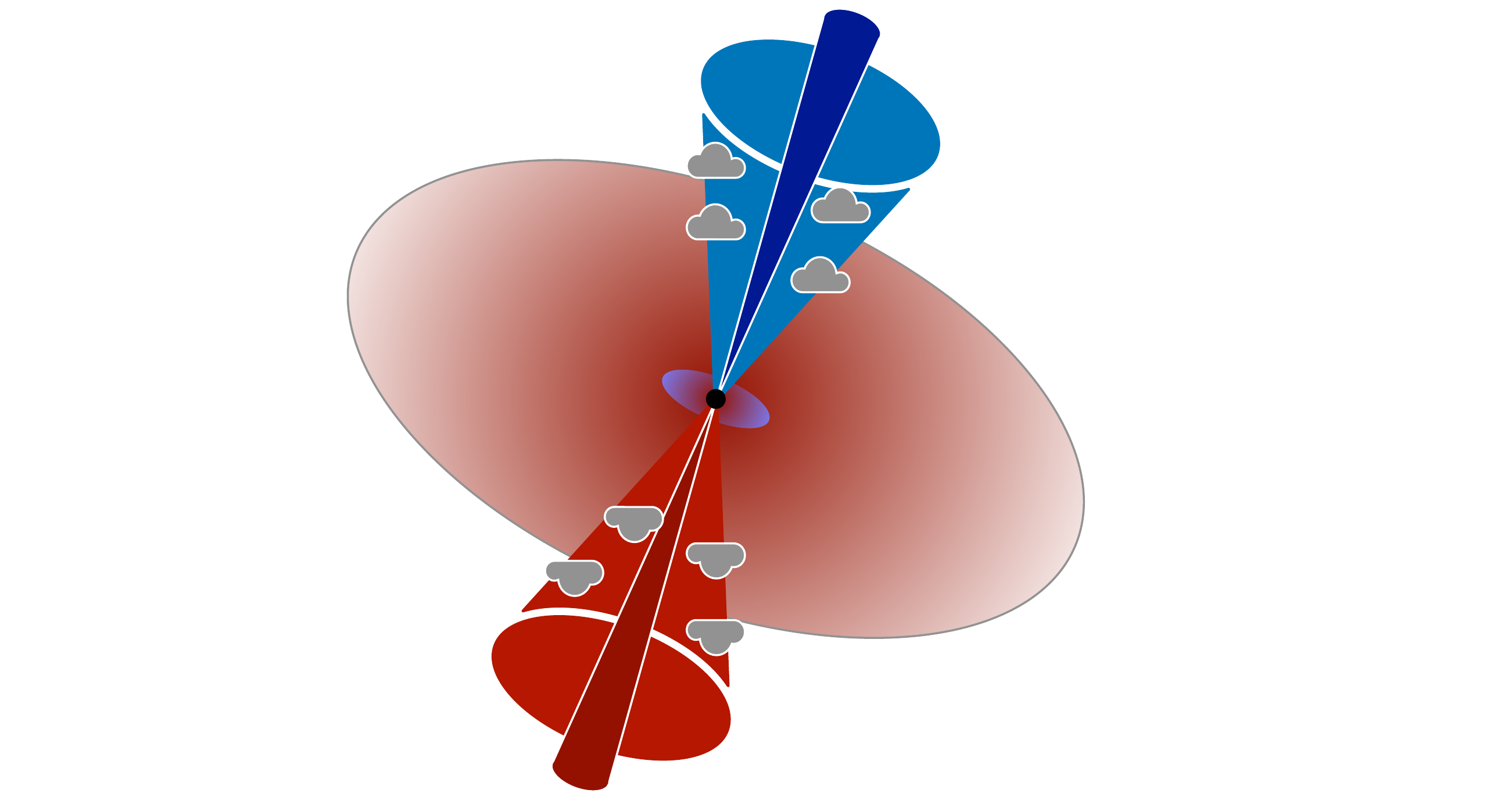}}
\caption{Cartoon showing a nested biconical structure in the outflow (that is blue-shifted when approaching and redshifted when receding as denoted by the colors) in order to reproduce the double-peaked emission line AGN spectroscopic observations. {The host galaxy (red ellipse) and accretion disk (blue ellipse) as shown here are not to scale. The~outer layers of a jet or a wind could be driving the NLR gas clouds.} }
\label{fig5}
\end{figure}

Furthermore, it was found that the double peaks of the emission lines were typically separated by velocities of $\sim$100--300~km~s$^{-1}$ for most sources, and~the widths of the lines corresponded to velocities of $\sim$100--200~km~s$^{-1}$ \citep{Kharb2015, Kharb2017a, Kharb2019, Kharb2021}. These velocities are much smaller, by~factors of several hundred, than the expected jet speeds. This, therefore, suggests that the emission line gas could be pushed in a direction lateral to the jet (e.g.,~\citep{Kharb2017a, Girdhar2022}) or could arise in wider, slower-moving winds around the jets (see for example Figure~\ref{fig5}). Indeed, nested biconical outflows have been invoked to explain the origin of double-peaked emission lines in low luminosity AGN by \citet{Nevin16, Nevin18}. Importantly, wide wind-like outflows can be efficient agents of ``AGN feedback'' (see the review by \citep{Harrison2018}). Indeed, from~the MAPPINGS III modeling of emission lines such as H$\alpha$, H$\beta$, H$\gamma$, [S II], \mbox{[O III],} and~[O II], it appears that in sources possessing  parsec-scale jets, the~``shock + precursor'' model can explain the observed line ratios, consistent with the idea of jet--NLR interaction. The~``shock + precursor'' model comes into play when ionizing radiation (i.e., extreme UV and soft X-ray photons) generated by the cooling of hot gas behind a shock front creates a strong radiation field leading to significant photoionization \citep{Allen2008}. The~spatial scales sampled by the SDSS optical fiber with a diameter of 3 arcsecs, corresponding to $\sim$3--6~kpc at the distance of the KISSR sources, are much larger than the $\sim$100 parsec-scale VLBA jets. However, as~\citet{Schmitt03a, Schmitt03b} have noted, the~NLR ranges from a few 100 parsecs to a few kpcs in Seyfert galaxies. The~emission-line modeling is consistent with the idea that RQ AGN are energetically capable of influencing their parsec and kpc-scale environments, making them agents of ``radio AGN~feedback''.
\section{Summary}\label{sec6}
{Radio-quiet AGN make up the vast majority of all AGN. While they lack the 100~kpc radio jets and lobes that are observed in the more spectacular looking and rarer radio-loud AGN, the~smaller outflows in RQ AGN have profound though subtle effects on their host galaxies. Multi-frequency polarization-sensitive observations of the PG RQ quasars and other Seyfert galaxies with the VLA and GMRT indicate that the radio outflows are layered or stratified. The~multi-component outflows could either be spine + sheath structures or jets with winds. While the ``sheath'' layers around the ``spines'' of jets could come about due to jet--medium interaction, they continue to entrain the surrounding gas, create instabilities between layers, and~further impact both the jet and the medium itself. Small (arcsec) scale jet--medium interaction is also implied by their polarimetric and jet kinetic power data. The~picture of nested biconical outflows also emerges from a completely different study, namely, the~VLBI study of a sample of Seyfert and LINER galaxies that exhibit double-peaked emission lines in their SDSS optical spectra. Two separate results point to jet--medium interaction in these sources. First, the~high incidence of one-sided radio jets, several of which are $\sim$100~parsec in extent, points to a selection bias when choosing a DPAGN sample as the double-peaked emission lines appear to be the result of jet--NLR gas cloud interaction. Second, plasma modeling codes on the optical emission lines from SDSS indicate that the NLR gas clouds are affected through the ``shock + precursor'' mechanism and~that the jets are likely stratified with a faster moving ``spine'' and a slower moving ``sheath'' or wind. Finally, multi-frequency arcsec-scale observations detect the signatures of multiple jet episodes in RQ AGN through the presence of additional steep-spectrum radio lobes. Episodic AGN activity in fact appears to be the norm in RQ AGN, signatures of which are seen in the radio spectra as well as morphological features like bow-shocks. In~the case of the Seyfert galaxies NGC2639 and NGC4051, and~others, these multiple jet episodes appear to excavate holes in the molecular gas in the central regions of their host galaxies, consistent with ``negative AGN feedback''. However, in~many cases, like the PG RQ quasars, there appears to be no currently observable depletion of molecular gas in their host galaxies. ``AGN feedback'', therefore, appears to be a complex phenomenon in the case of RQ AGN. It cannot be ruled out on small spatial scales as implied by the signatures of jet--medium interaction. However, global impact signatures on the host galaxies are often difficult to see. }

\vspace{6pt}

\authorcontributions{This paper has contributions from projects lead by both P.K. and S.S. Projects lead by S.S. are a part of her Ph.D. thesis currently being supervised by P.K. All authors have read and agreed to the published version of the manuscript. }

\funding{This research received no external funding.}

\informedconsent{Informed consent was obtained from all subjects involved in the~study.}

\dataavailability{The data underlying this article will be shared on reasonable request to the corresponding author.}

\acknowledgments{{We thank the anonymous reviewers for their suggestions that have significantly improved this~paper.} We thank the staff of the GMRT that made these observations possible. GMRT is run by the National Center for Radio Astrophysics of the Tata Institute of Fundamental Research. P.K. and S.S. acknowledge the support of the Department of Atomic Energy, Government of India, under~the project 12-R\&D-TFR-5.02-0700. The~National Radio Astronomy Observatory is a facility of the National Science Foundation operated under cooperative agreement by Associated Universities,~Inc.}

\conflictsofinterest{The authors declare no conflict of~interest.}\clearpage

\begin{adjustwidth}{-\extralength}{0cm}

%
%
\printendnotes[custom]

\reftitle{Referen{ces}}

%
%
%
%

%
%
%
%
%
%
%
%
%
%
%

%
%
%
%
%
%
%
%
%

%
%
%

%
%
%
%
%
%
%
%
\PublishersNote{}
\end{adjustwidth}
\end{document}